\documentclass[conference]{IEEEtran}
\IEEEoverridecommandlockouts
\usepackage{cite}
\usepackage{amsmath,amssymb,amsfonts}
\usepackage{algorithmic}
\usepackage{graphicx}
\usepackage{textcomp}
\usepackage{xcolor}
\usepackage{url}
\usepackage{tcolorbox}
\usepackage{multirow}
\usepackage{tcolorbox}
\usepackage{setspace}

\def\BibTeX{{\rm B\kern-.05em{\sc i\kern-.025em b}\kern-.08em
    T\kern-.1667em\lower.7ex\hbox{E}\kern-.125emX}}

\begin{document}

\title{A Comparison of Vulnerability Feature Extraction Methods from Textual Attack Patterns}

\author{\IEEEauthorblockN{Refat Othman}
\IEEEauthorblockA{\textit{Faculty of Engineering} \\
\textit{Free University of Bozen-Bolzano}\\
Bolzano, Italy \\
ramneh@unibz.it}
\and
\IEEEauthorblockN{Bruno Rossi}
\IEEEauthorblockA{\textit{Faculty of Informatics} \\
\textit{Masaryk University}\\
Brno, Czech Republic \\
brossi@mail.muni.cz}
\and

\IEEEauthorblockN{Barbara Russo}
\IEEEauthorblockA{\textit{Faculty of Engineering} \\
\textit{Free University of Bozen-Bolzano}\\
Bolzano, Italy \\
brusso@unibz.it}

}

\maketitle

\begin{abstract}

Nowadays, threat reports from cybersecurity vendors incorporate detailed descriptions of attacks within unstructured text.
Knowing vulnerabilities that are related to these reports helps cybersecurity researchers and practitioners understand and adjust to evolving attacks and develop mitigation plans. This paper aims to aid cybersecurity researchers and practitioners in choosing attack extraction methods to enhance the monitoring and sharing of threat intelligence. In this work, we examine five feature extraction methods (TF-IDF, LSI, BERT, MiniLM, RoBERTa) and find that Term Frequency-Inverse Document Frequency (TF-IDF) outperforms the other four methods with a precision of 75\% and an F1 score of 64\%. The findings offer valuable insights to the cybersecurity community, and our research can aid cybersecurity researchers in evaluating and comparing the effectiveness of upcoming extraction methods.
\end{abstract}


\enlargethispage{\baselineskip}

\section{Introduction}
Cyberattacks are causing significant financial harm, with the cost reached \$6 trillion in 2022 and are on track to reach \$10.5 trillion by 2025~\cite{Cybercrime}. The rapidly evolving threat landscape makes thwarting cyberattacks a bigger challenge. Hence, cyber threat intelligence (CTI) sharing and ongoing monitoring have become the highest priorities. Cybersecurity vendors publish CTI reports describing how attackers utilize the techniques and which techniques and patterns are used for performing the attack~\cite{sun2023cyber}. Additionally, employing CTI enables companies to take a proactive and prevent possible attacks before they have a chance to do any damage~\cite{elder2022really}~\cite{sakellariou2022reference}.
In this context, methods are necessary to extract the vulnerability from the description of the attack. Thus, knowing the vulnerability of the attacks can aid cybersecurity practitioners in developing detection and mitigation strategies for attacks~\cite{sakellariou2022reference}~\cite{rahman2022threat}. 
The Common Vulnerabilities and Exposures (CVE) list identifies vulnerabilities in the computational logic of hardware and software components that, if exploited, might compromise availability, confidentiality, or integrity~\cite{WhatCVE}~\cite{refat2024cybersecurity}. Moreover, Common Weakness Enumeration (CWE) is a community-developed collection of typical weaknesses in software, coding errors, and security flaws~\cite{CWE}.

Common Attack Patterns Enumeration and Classification (CAPEC) is a complete dictionary of known attack patterns and weaknesses, including IoT devices, hardware appliances, and software applications~\cite{CAPEC}. In addition, the attack pattern contains a large amount of text, and manually extracting vulnerabilities 
is a crucial step for utilizing attack patterns effectively. It is a time-consuming and error-prone task due to the extensive amount of text. Thus, cybersecurity teams have proposed automated feature extraction methods from attack reports, including Term Frequency-Inverse Document Frequency (TF-IDF), Latent Semantic Indexing (LSI), Deep Self-Attention Distillation for Task-Agnostic Compression of Pre-Trained Transformers (MiniLm), a Robustly Optimized BERT Pretraining Approach (RoBERTa), and Bidirectional Encoder Representations from Transformers (BERT).  These methods use Natural Language Processing (NLP) and Machine Learning (ML) methods to classify texts. Therefore, a comparison study of these methods would provide the best method for extracting the vulnerability of attack patterns. 
In this work, we evaluate and compare the performance of these five methods in classifying text, specifically attack pattern descriptions, to their corresponding CVEs.
\textit{ The goal of this paper is to aid cybersecurity researchers in choosing attack extraction methods for extracting vulnerability information, allowing them to prioritize actions, strengthen defenses, and stay ahead of the evolving cyber threat landscape.} Our study's dataset and source code can be downloaded from Github~\cite{VULDAP}. Thus, we aim to answer the following question.
\enlargethispage{\baselineskip}

\par\noindent

     \textit{\textbf{RQ: } How do different feature extraction methods compare in terms of performance when classifying textual descriptions of attack patterns to CVE issues across different classifiers?} 

We include a list of our contributions: 
\begin{itemize}
    \item A comparison analysis of the five methods for feature extraction;
    \item A sensitivity analysis examining how applying multiclass classification affects the methods under comparison;
    \item A novel mapping dataset~\cite{VULDAP} explicitly links attack patterns with vulnerabilities found in MITRE repositories;
\end{itemize}


This paper is structured as follows. Section~\ref{Sec:METHODOLOGY} outlines the methodology we propose to use for our pipeline comparison. We summarise our preliminary results in section~\ref{Sec:results}. Section~\ref{Sec:limitation} outlines the limitations of our work. Finally, Section~\ref{Sec:conclusion} concludes with our results and future research.

\section{Study Design and Methodology }
\label{Sec:METHODOLOGY}

In this section, we describe our research methodology for comparing vulnerability feature extraction methods. Fig~\ref{fig:OverviewMethidology} illustrates the implemented pipeline. 
\begingroup
\setlength{\abovedisplayskip}{0pt}
\setlength{\belowdisplayskip}{0pt}
\begin{figure}[hbt]
    \centering
    \includegraphics[width=\linewidth]{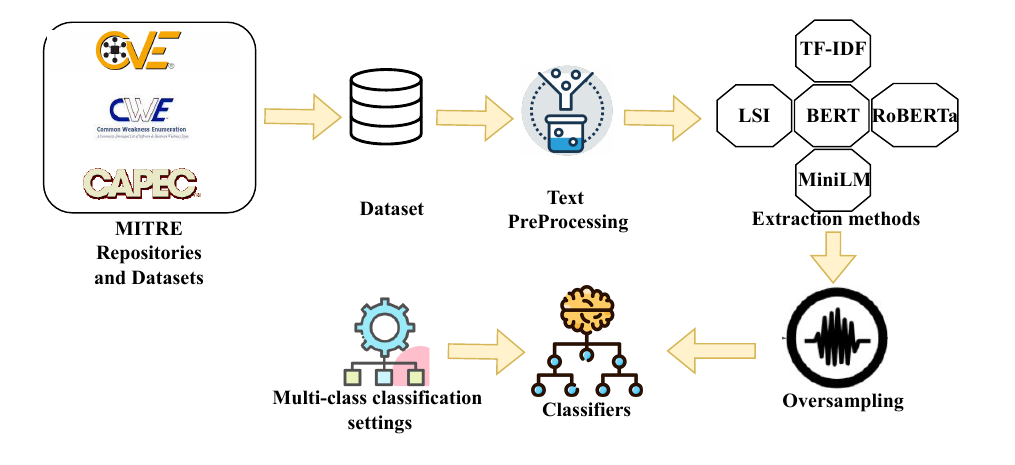}
    \caption{Methodology Overview}\label{fig:OverviewMethidology}
\end{figure}

\endgroup\ignorespaces


\par\noindent
\subsection{Dataset Collection}
 \label{sec:vuldataset}
\enlargethispage{\baselineskip}

We construct and collect our dataset from MITRE~\cite{MITRE} repositories and datasets, CAPEC (Attack Pattern) ~\cite{CAPEC}, CWE~\cite{CWE}, and CVE ~\cite{CVEdataset} as illustrated in Table~\ref{tab:golddataset} and Fig~\ref{fig:MitreConnection}.
We build the dataset by utilizing the two pairings of related links: attack patterns are linked to weaknesses, and weaknesses are linked to vulnerabilities. Links in the dataset make it easier to go from  attack pattern to vulnerability  (or vice versa). The summary of the mapping is as follows: (1) Connecting attack pattern and CWE:
Attack patterns and CWE reports are connected through the CWE-ID.
By linking CWE with attack patterns, the dataset enables a clear understanding of which specific attack patterns target weaknesses. (2) Connecting CWE and CVE:
CWE and CVE reports are connected through the CVE-ID. This linkage allows for the association of specific vulnerabilities with broader categories of weaknesses they exploit. By leveraging this linkage, cybersecurity professionals can gain a more comprehensive understanding of the underlying weaknesses that lead to the emergence of particular vulnerabilities. The dataset combines data from various sources to give users a comprehensive view of vulnerabilities, their underlying weaknesses, and their potential impact. We found that only 133 attack patterns are linked to 106 CWE reports, and the same attack patterns are connected to 685 CVE reports.



\begingroup
\setlength{\abovedisplayskip}{0pt}
\setlength{\belowdisplayskip}{0pt}

{\fontsize{6pt}{6pt}\selectfont

 \begin{table}[htb]
\caption{attack pattern descriptions linked and not linked to vulnerability reports\label{tab:golddataset}}
\centering
\bgroup
\def\arraystretch{1}
\setlength{\tabcolsep}{4pt}
\small
\begin{tabular}{lccc}
\hline  
&\textbf{Linked} &\textbf{Not linked} & \textbf{Total} \\
\hline
\textbf{Attack Patterns~\cite{CAPEC}} & 143& 416& 559 \\
\textbf{CWE reports~\cite{CWE}} &149 &786 &935\\
\textbf{CVE reports~\cite{CVEdataset}} &685 & 294919&295604\\
\hline
\end{tabular}%
\egroup
\end{table}
}
\endgroup\unskip


\begingroup
\setlength{\abovedisplayskip}{0pt}
\setlength{\belowdisplayskip}{0pt}
\begin{figure}[htb]
\centering
\small
\includegraphics[width=0.6\columnwidth]{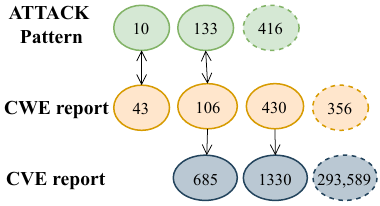}
\caption{MITRE repositories and their connections}\label{fig:MitreConnection}
\end{figure}
\endgroup\unskip

\par\noindent
\subsection{Text Pre-Processing}

Fig \ref{fig:textpreprocessing} illustrates the steps of pre-processing.
The first step is to convert all descriptions to lowercase. Following this, we proceed to remove spaces, punctuation, and stop words, which are commonly occurring terms typically excluded from analysis. Then, perform stemming, tokenization, and lemmatization using the \texttt{gensim} python library and \texttt{preprocessing} function.
Tokenization is the process of splitting unstructured text input into tokens. Stemming involves the process of reducing words to their base or root form. Lemmatization employs a dictionary
search to determine a word’s precise form based on its
part of speech.  


\begingroup
\setlength{\abovedisplayskip}{0pt}
\setlength{\belowdisplayskip}{0pt}
\begin{figure}[htb]
\centering
\includegraphics[width=.9\columnwidth]{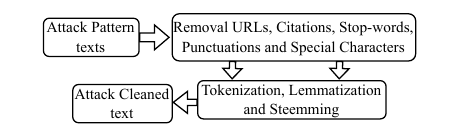}
\caption{Text pre-processing\label{fig:textpreprocessing}}
\end{figure}
\endgroup
\unskip


\par\noindent
\subsection{Feature Extraction}
\label{sec:FeatureExtraction}
This section described the employed methods as follows:
\par\noindent
\textbf{TF-IDF} method is the widely used approach for vectorizing text documents ~\cite{rahman2022threat}~\cite{othman2023vuldat}. We utilize \texttt{TfidfVectorizer} from the \texttt{scikit-learn} python package to conduct feature extraction. Initially, we compute the vectors of the corpus. Following this, we normalize the TF-IDF vectors to ensure they have the same length. Finally, these normalized TF-IDF vectors are inputted into classifiers to evaluate the method.
\textbf{LSI} is a mathematical model that examines connections between a group of documents and the words present within them~\cite{rahman2022threat}. 
We utilize \texttt{lsimodel} and \texttt{tfidfmodel} from the \texttt{scikit-learn} python package to conduct feature extraction. 
\textbf{BERT}~\cite{reimers2019sentence} is a deep learning model introduced by Google in 2018. 
We utilize \texttt{bert-base-uncased} from the \texttt{sentence\_transformers} python package to conduct feature extraction. 
\textbf{MiniLM}, the paraphrase-multilingual-MiniLM-L12-v2 model~\cite{paraphrasemini} is a pre-trained sentence transformer model that utilizes the MiniLM architecture with 12 layers.
To extract features, we use \texttt{paraphrase-multilingual-MiniLM} from the \texttt{sentence\_transformers} python package. 
\textbf{RoBERTa}~\cite{liu2019roberta} is a new version of the BERT model developed by Facebook AI that includes removing the next sentence prediction objective, longer training periods with more data, and dynamic masking throughout training. 
We utilize \texttt{roberta-base} from the \texttt{sentence\_transformers} python package for feature extraction. 

\enlargethispage{\baselineskip}

\par\noindent
\subsection{Oversampling}
The attack patterns do not have the same number of CVE issues in our dataset. To address the issue of imbalanced classification, we employ the Synthetic Minority Oversampling Technique (\texttt{SMOTE}) method for oversampling to mitigate imbalanced classification problems. We used SMOTE on the computed features in each method because it can only be used for the oversampling of numerical features. Thus, this approach helps mitigate the imbalanced distribution within our dataset.
\enlargethispage{\baselineskip}


\par\noindent
\subsection{Classification}
In this study, we employ six classifiers: \texttt{Random Forest} (RF), \texttt{K-nearest Neighbor} (KNN), \texttt{Neural Network} (NN), \texttt{Naive Bayes} (NB), and \texttt{Support Vector Machine} (SVM) classifiers. We used \texttt{RandomForestClassifier, KNeighborsClassifier, MLPClassifier,  GaussianNB, DecisionTreeClassifier} and \texttt{SVC}  for the classifiers using algorithms available in the \texttt{scikit-learn} python package.



\par\noindent
\subsection{Classification Settings and Cross Validation}

Each instance of an attack pattern in our dataset can be linked to one of up to 685 distinct CVE issues, making this experiment a multiclass classification task. To evaluate the classifiers, we apply the classifiers to each method in the following two cases: 
\begin{itemize}
    \item \textbf{Binary Classification:} An attack pattern is classified into one of two possible CVE issues. Here, we rank the CVE issues based on the frequency of their corresponding attack pattern descriptions within the dataset. We then select the top two CVE issues and their associated descriptions for classification.
    \item \textbf{Multiclass Classification:} An attack pattern is classified into one of multiple CVE issues. We again rank the CVE issues by the number of corresponding attack pattern descriptions and choose the top $n$ CVE issues,
    where $n$ is a variable that represents various classification scenarios.
    We test each method across five different scenarios with $n = 2, 3, 4, 5, 6$ to evaluate the classifiers' performance under varying complexities.
\end{itemize}
We divide our dataset into distinct training, validation, and testing sets using a modified K-fold cross-validation technique with $K=5$. For each fold, we use 80\% of the dataset as the training set, while 10\% for testing and 10\% for validation. This approach enhances the performance of our models by training and evaluating them across different subsets of the data.


\par\noindent
\section{Preliminary Results}
\label{Sec:results}

\subsection{RQ: How do different feature extraction methods compare in terms of performance when classifying textual descriptions of attack patterns to CVE issues across different classifiers?}
To answer this RQ, we evaluated the performance of five feature extraction methods in extracting the vulnerability from attack pattern descriptions. 
Table~\ref{tab:performance} illustrates the performance of all methods with all classifiers, including precision, recall, F1, and AUC. 
\begin{table}[h]
\centering
\begin{scriptsize}
\caption{Performance of methods across all multiclass classification settings (unit is a percentage)}\label{tab:performance}
\begin{tabular}{p{,15cm}|p{0.55cm}|*{4}{p{1.3cm}}}
\hline
\rotatebox[origin=c]{90}{\textbf{Method}} & \rotatebox[origin=c]{90}{\textbf{Classifier}} & \textbf{Precision} & \textbf{Recall} & \textbf{F1 Score} & \textbf{AUC} \\
\hline
 \multirow{5}{*}{\rotatebox[origin=c]{90}{\textbf{TF-IDF}}} & KNN & 43-92(62) & 41-92(57) & 38-92(54) & 67-94(77) \\ 
 & NB & \textbf{59-91(75)} & \textbf{53-86(66)} & \textbf{49-85(64)} & 76-86(81) \\ 
 & SVM & 50-94(70) & 50-90(65) & \textbf{46-88(64)} & \textbf{74-99(85)} \\ 
 & RF & 55-93(69) & 50-90(64) & \textbf{49-90(64)} & 76-99(84) \\ 
 & DT & 46-92(66) & 41-89(59) & 40-89(59) & 71-89(77) \\ 
 & NN & 53-94(69) & 49-91(64) & 47-91(63) & \textbf{79-98(85)} \\ 
 \hline
 \multirow{5}{*}{\rotatebox[origin=c]{90}{\textbf{LSI}}} & KNN & 44-93(62) & 41-92(57) & 38-92(55) & 65-92(77) \\ 
 & NB & 35-77(51) & 39-72(50) & 33-70(44) & 73-80(76) \\ 
 & SVM & \textbf{50-93(69)} & \textbf{50-89(64)} & \textbf{46-87(63)} & \textbf{76-98(85)} \\ 
 & RF & 47-94(64) & 45-92(61) & 43-91(60) & 72-93(81) \\ 
 & DT & 45-93(64) & 42-91(58) & 40-90(57) & 68-91(76) \\ 
 & NN & 46-64(56) & 44-59(53) & 41-59(52) & 76-83(79) \\ 
 \hline
\multirow{5}{*}{\rotatebox[origin=c]{90}{\textbf{MiniLM}}}  & KNN & 49-92(66) & 43-90(60) & 41-89(58) & 74-92(80) \\ 
 & NB & 48-92(65) & 44-89(61) & 43-89(60) & 75-93(82) \\ 
 & SVM & \textbf{47-92(69)} & 48-88(63) & 43-87(61) & 75-99(85) \\ 
 & RF & 52-90(66) & 48-84(61) & 46-82(60) & 74-97(82) \\ 
 & DT & 46-86(65) & 44-83(60) & 42-82(59) & 73-83(77) \\ 
 & NN & 53-92(68) & \textbf{49-88(64)} &\textbf{48-87(63)} & \textbf{79-98(86)} \\ 
 \hline
\multirow{5}{*}{\rotatebox[origin=c]{90}{\textbf{RoBERTa}}}  & KNN & 48-77(60) & 44-76(55) & 41-76(53) & 67-85(75) \\ 
 & NB & 51-89(65) & 48-86(61) & 47-85(60) & 76-89(82) \\ 
 & SVM & 15-46(28) & 28-55(38) & 16-42(26) & 43-69(59) \\ 
 & RF & 52-90(67) & 47-85(62) & 46-84(61) & 74-98(83) \\ 
 & DT & 45-89(65) & 45-87(61) & 42-86(60) & 72-87(78) \\ 
 & NN & \textbf{51-92(68)} & \textbf{50-90(64}) & \textbf{47-89(63)} & \textbf{78-97(85)} \\ 
 \hline
 \multirow{5}{*}{\rotatebox[origin=c]{90}{\textbf{BERT}}} & KNN & 48-70(56) & 45-68(52) & 41-67(50) & 67-80(74) \\ 
 & NB & 50-86(64) & 45-82(58) & 44-81(57) & 73-86(79) \\ 
 & SVM & \textbf{49-92(69)} & 45-91(61) & 41-90(60) & \textbf{77-97(85)} \\ 
 & RF & 51-90(67) & 47-85(61) & 46-84(60) & 74-95(81) \\ 
 & DT & 43-78(62) & 41-71(55) & 39-69(54) & 69-75(73) \\ 
 & NN & 49-92(68) &\textbf{47-89(63)} & \textbf{45-88(61)} & \textbf{78-98(85)}
\\
\hline
\end{tabular}
\end{scriptsize}

\end{table}
\enlargethispage{\baselineskip}
Every corresponding cell in the table shows the score in the format \text{A-B(C)}, where A represents the lowest observed score, B is the highest observed score, and C is the arithmetic average score. For example,  the top right cell contains 67-94(77), which is the KNN classifier paired with the AUC score about the TF-IDF method. The lowest, highest, and average scores from all possible classification settings $(n=2, 3, 4, 5, 6)$ are 67, 94, and 77, respectively.
The bold cells display the maximum average scores for each method when paired with six classifiers.
\par\noindent
\begin{figure}[h]
    \centering
    \includegraphics[width=\columnwidth]{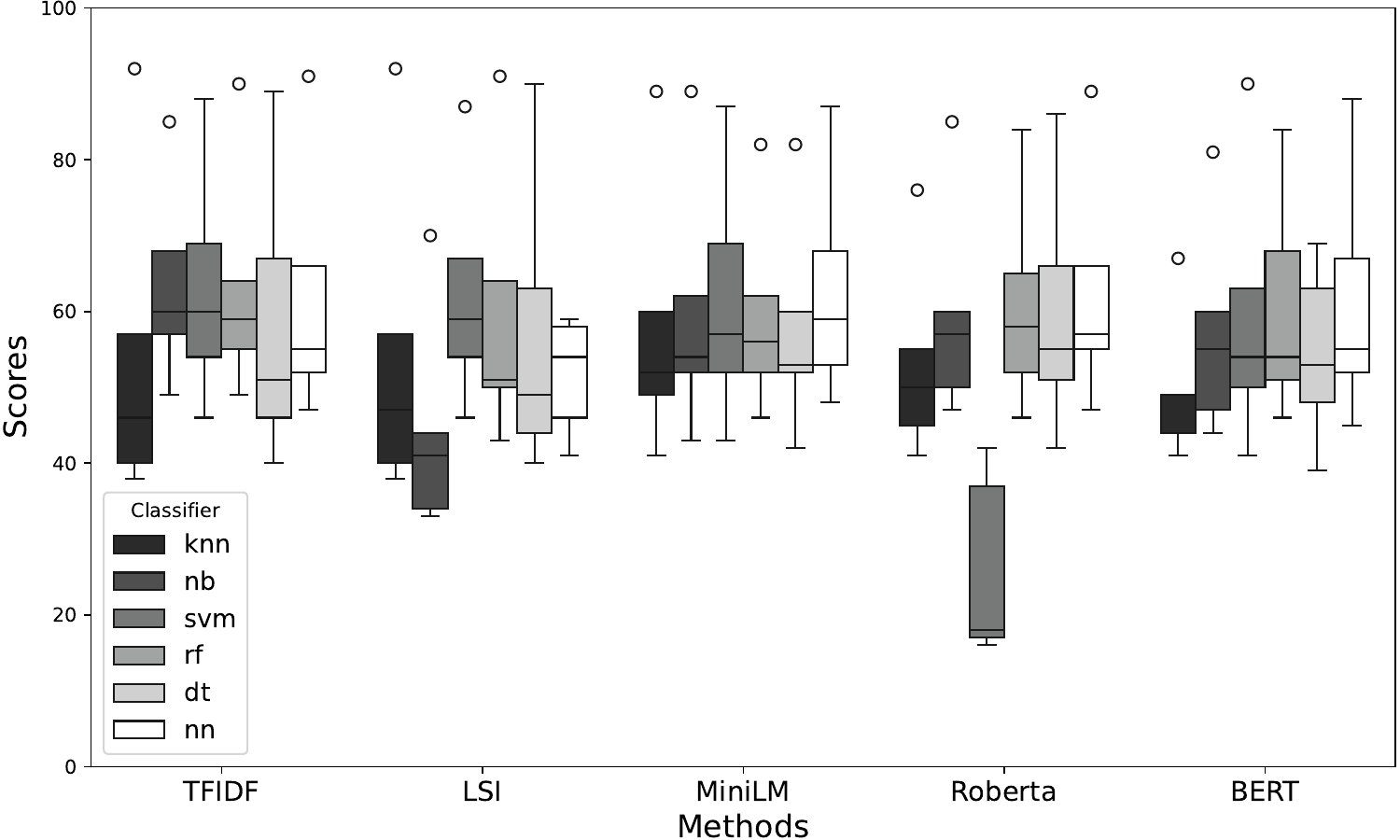}
    \caption{Boxplot of F1 score for five methods}
    \label{fig:f1withclassifiers}

\end{figure}
\begin{figure}[h]
    \centering
    \includegraphics[width=\columnwidth]{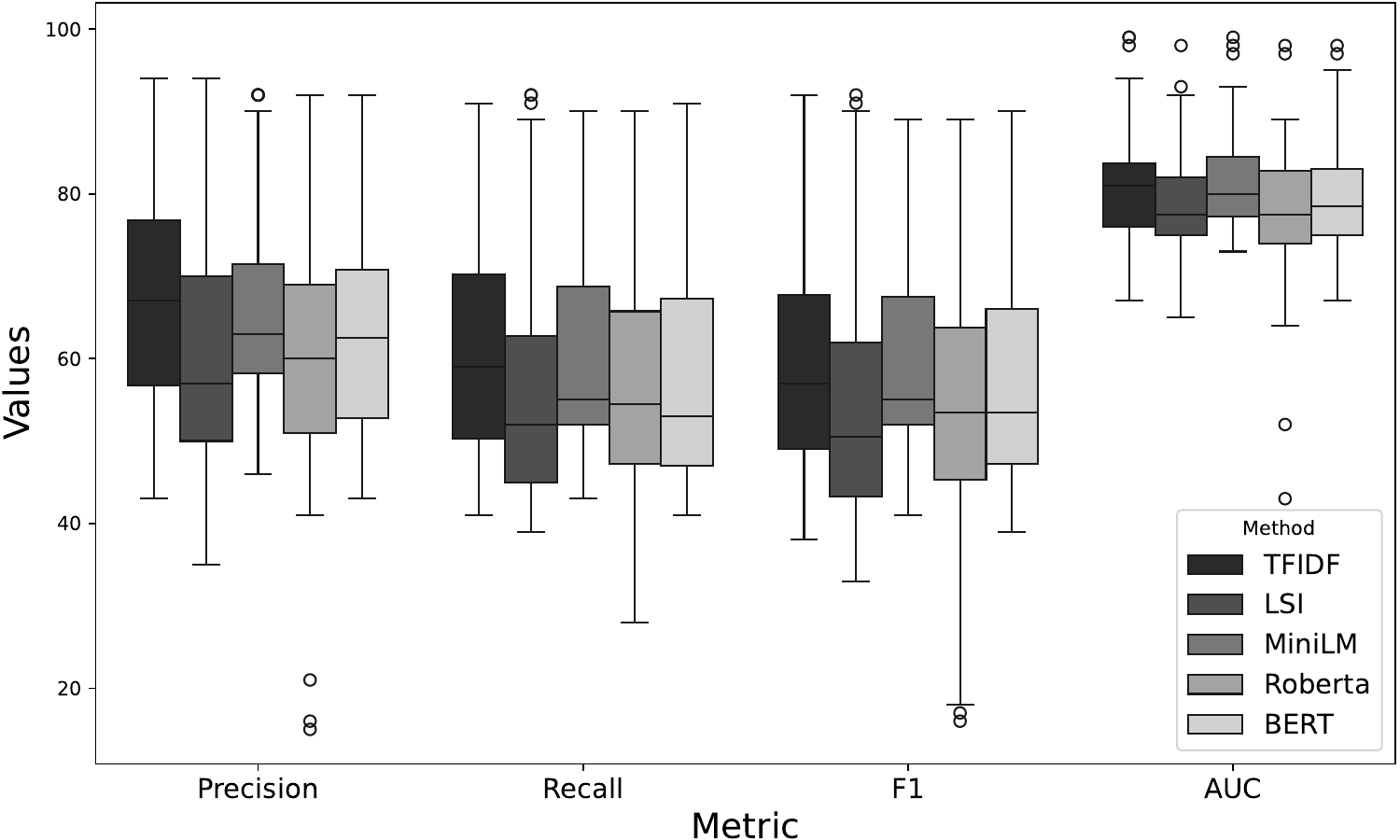}
    \caption{Boxplot of five methods on performance metrics}
    \label{fig:PerformanceM}

\end{figure}

\enlargethispage{\baselineskip}
Fig~\ref{fig:f1withclassifiers} presents the F1 score boxplot for all methods, while Fig~\ref{fig:PerformanceM} presents all performance metrics for the methods. The following are our findings: NB, SVM, and NN work best for the TF-IDF method. We find that NB shows the best performance in precision (75) and F1 Score (64), while SVM shows the best in F1 score (64), and SVM and NN both show the best in AUC (85) as shown in Table~\ref{tab:performance}. Out of the six classifiers, KNN performs the worst for all performance scores.  SVM works best for the LSI method. We find SVM classifiers perform the best in all performance scores. SVM and NN work best for the MiniLM method. We find that the SVM classifier performs the best performance in precision (69). 
\enlargethispage{\baselineskip}
The NN classifier shows the best performance in recall (64), F1 score (63), and AUC (86). The SVM classifier differs by 1\% in precision score compared with the NN classifier. In addition, we find that, for the MiniLM method, all six classifiers obtain nearly identical scores, with differences between them not exceeding 5\%. 

SVM and NN work best for the BERT method, with SVM achieving the highest precision (69) and AUC (85), while NN excels in recall (63), F1 score (61), and AUC (85). The SVM classifier differs by 1\%, 2\%, and 1\% in precision, recall, and F1 scores compared to NN.
\enlargethispage{\baselineskip}

\enlargethispage{\baselineskip}

\section{Threats to Validity}
\label{Sec:limitation}

In this section, we provide the limitations of this study. We did not compare the methods using a large number of reports, such as the descriptions of CVE and attack patterns.  Additionally, we assume that one attack report corresponds to a single corresponding CVE. However, attack patterns may have multiple corresponding CVEs, which multi-label classifiers can classify. Furthermore,  for classification tasks, the dataset contains 685 CVEs, and the classification performance of the remaining links is not evaluated. 


\section{Related Work}
\label{Sec:relatedwork}
Cyber threat intelligence enables companies to take a proactive approach to cybersecurity, enabling them to prepare for and prevent possible attacks before they can do any damage ~\cite{sun2023cyber}~\cite{sakellariou2022reference}. 
The author in~\cite{rahman2022threat} paper compares various attack technique extraction methods derived from threat reports to aid cybersecurity researchers and practitioners in monitoring and sharing CTI. The findings reveal that two methods, leveraging TFIDF and LSI, demonstrate superior performance compared to the other three methods, BM25, TFIDF-NP, and LSI-Co, as indicated by their higher F1 scores.
The Cve2att\&ck~\cite{grigorescu2022cve2att} is a model that utilizes BERT-based language models to associate CVE automatic reports with techniques based on the text description found in CVE metadata. The CVE Transformer (CVET)~\cite{ampel2021linking} model combines the benefits of utilizing RoBERTa~\cite{liu2019roberta} to link CVEs to ten ATT\&CK Enterprise Matrix tactics.

\section{Conclusion}
\label{Sec:conclusion}

This study compared five methods (TF-IDF, LSI, MiniLM, RoBERTa, BERT) to classify text descriptions linking attack patterns to CVE vulnerabilities. TF-IDF achieved the best results with 75\% precision and 64\% F1 score. We recommend cybersecurity researchers choose the best features for text-to-CVE extraction. We plan to explore more feature extraction methods and share their findings to address the missing links in MITRE's repositories. Additionally, we will fine-tune the classifiers beyond their default parameters to optimize classification performance.
\enlargethispage{\baselineskip}
\section{Acknowledgements}
The first author thanks the CSLab at the Free University of Bozen-Bolzano for supporting this work under project no. EFRE1039 in the EFRE-FESR 2021-2027 program.
\enlargethispage{\baselineskip}
\enlargethispage{\baselineskip}

\bibliographystyle{unsrt}
\bibliography{sample-base}

\begin{thebibliography}{10}

\bibitem{Cybercrime}
Mac~Margolis Robert~Muggah.
\newblock Cybercrime to cost the world 10.5 trillion annually by 2025, 2023.
\newblock Accessed: January 28, 2024. \url{ https://www.weforum.org/agenda/2023/01/global-rules-crack-down-cybercrime/}.

\bibitem{sun2023cyber}
Nan Sun, Ming Ding, Jiaojiao Jiang, Weikang Xu, Xiaoxing Mo, Yonghang Tai, and Jun Zhang.
\newblock Cyber threat intelligence mining for proactive cybersecurity defense: A survey and new perspectives.
\newblock {\em IEEE Communications Surveys \& Tutorials}, 2023.

\bibitem{elder2022really}
Sarah Elder, Nusrat Zahan, Rui Shu, Monica Metro, Valeri Kozarev, Tim Menzies, and Laurie Williams.
\newblock Do i really need all this work to find vulnerabilities? an empirical case study comparing vulnerability detection techniques on a java application.
\newblock {\em Empirical Software Engineering}, 27(6):154, 2022.

\bibitem{sakellariou2022reference}
Georgios Sakellariou, Panagiotis Fouliras, Ioannis Mavridis, and Panagiotis Sarigiannidis.
\newblock A reference model for cyber threat intelligence (cti) systems.
\newblock {\em Electronics}, 11(9):1401, 2022.

\bibitem{rahman2022threat}
Md~Rayhanur Rahman and Laurie Williams.
\newblock From threat reports to continuous threat intelligence: A comparison of attack technique extraction methods from textual artifacts.
\newblock {\em arXiv preprint arXiv:2210.02601}, 2022.

\bibitem{WhatCVE}
Taylor Armerding.
\newblock Cve definitions, 2017.
\newblock Accessed: May 4, 2023. \url{https://www.csoonline.com/article/3204884/what-is-cve-its-definition-and-purpose.html}.

\bibitem{refat2024cybersecurity}
Othman Refat, Rossi Bruno, and Russo Barbara.
\newblock Cybersecurity defenses: Exploration of cve types throug attack descriptions.
\newblock In {\em 2024 50th Euromicro Conference on Software Engineering and Advanced Applications (SEAA)}. IEEE, 2024.

\bibitem{CWE}
MITRE.
\newblock Cwe dataset, 2024.
\newblock \url{https://cwe.mitre.org/}.

\bibitem{CAPEC}
MITRE.
\newblock Capec, 2023.
\newblock \url{https://capec.mitre.org/}.

\bibitem{VULDAP}
Code and dataset, 2024.
\newblock Accessed: Feb 22, 2024. \url{https://github.com/ref3t/VulnerabilityExtractionMethods}.

\bibitem{MITRE}
MITRE.
\newblock Mitre att\&ck, 2024.
\newblock Accessed: January 8, 2023. \url{https://attack.mitre.org/}.

\bibitem{CVEdataset}
MITRE.
\newblock Cve, 2023.
\newblock \url{https://cve.mitre.org/}.

\bibitem{othman2023vuldat}
Refat Othman and Barbara Russo.
\newblock Vuldat: Automated vulnerability detection from cyberattack text.
\newblock In {\em International Conference on Embedded Computer Systems}, pages 494--501. Springer, 2023.

\bibitem{reimers2019sentence}
Nils Reimers and Iryna Gurevych.
\newblock Sentence-bert: Sentence embeddings using siamese bert-networks.
\newblock {\em arXiv preprint arXiv:1908.10084}, 2019.

\bibitem{paraphrasemini}
Hugging Face.
\newblock paraphrase-multilingual-minilm-l12-v2, 2024.
\newblock Accessed: July 2, 2023. \url{https://huggingface.co/sentence-transformers/paraphrase-multilingual-MiniLM-L12-v2}.

\bibitem{liu2019roberta}
Yinhan Liu, Myle Ott, Naman Goyal, Jingfei Du, Mandar Joshi, Danqi Chen, Omer Levy, Mike Lewis, Luke Zettlemoyer, and Veselin Stoyanov.
\newblock Roberta: A robustly optimized bert pretraining approach.
\newblock {\em arXiv preprint arXiv:1907.11692}, 2019.

\bibitem{grigorescu2022cve2att}
Octavian Grigorescu, Andreea Nica, Mihai Dascalu, and Razvan Rughinis.
\newblock Cve2att\&ck: Bert-based mapping of cves to mitre att\&ck techniques.
\newblock {\em Algorithms}, 15(9):314, 2022.

\bibitem{ampel2021linking}
Benjamin Ampel, Sagar Samtani, Steven Ullman, and Hsinchun Chen.
\newblock Linking common vulnerabilities and exposures to the mitre att\&ck framework: A self-distillation approach.
\newblock {\em arXiv preprint arXiv:2108.01696}, 2021.

\end{thebibliography}

\end{document}